\documentstyle[epsf]{mn}     

\newif\ifAMStwofonts

\ifoldfss 
  \newcommand{\rmn}[1] {{\rm #1}} 
  \newcommand{\itl}[1] {{\it #1}} 
  \newcommand{\bld}[1] {{\bf #1}} 
  \ifCUPmtlplainloaded \else 
    \NewTextAlphabet{textbfit} {cmbxti10} {} 
    \NewTextAlphabet{textbfss} {cmssbx10} {} 
    \NewMathAlphabet{mathbfit} {cmbxti10} {} 
    \NewMathAlphabet{mathbfss} {cmssbx10} {} 
  \fi 
  \ifAMStwofonts 
    \ifCUPmtlplainloaded \else
      \NewSymbolFont{upmath} {eurm10} 
      \NewSymbolFont{AMSa} {msam10} 
      \NewMathSymbol{\upi}     {0}{upmath}{19} 
      \NewMathSymbol{\umu}     {0}{upmath}{16} 
      \NewMathSymbol{\upartial}{0}{upmath}{40} 
      \NewMathSymbol{\leqslant}{3}{AMSa}{36} 
      \NewMathSymbol{\geqslant}{3}{AMSa}{3E} 
      \let\oldle=\le     \let\oldleq=\leq 
      \let\oldge=\ge     \let\oldgeq=\geq 
      \let\leq=\leqslant \let\le=\leqslant 
      \let\geq=\geqslant \let\ge=\geqslant 
    \fi 
  \fi 
\fi 
 
\ifnfssone 
  \newmathalphabet{\mathit} 
  \addtoversion{normal}{\mathit}{cmr}{m}{it} 
  \addtoversion{bold}{\mathit}{cmr}{bx}{it} 
  \newcommand{\rmn}[1] {\mathrm{#1}} 
  \newcommand{\itl}[1] {\mathit{#1}} 
  \newcommand{\bld}[1] {\mathbf{#1}} 
  \def\textbfit{\protect\txtbfit} 
  \def\textbfss{\protect\txtbfss} 
  \long\def\txtbfit#1{{\fontfamily{cmr}\fontseries{bx}\fontshape  
\long\def\txtbfit#1{{\fontfamily{cmr}\fontseries{bx}\fontshape{it}
    \selectfont #1}} 
  
\long\def\txtbfss#1{{\fontfamily{cmss}\fontseries{bx}\fontshape{n
}%
    \selectfont #1}} 
  \newmathalphabet{\mathbfit} 
\newmathalphabet{\mathbfit} 
  \addtoversion{normal}{\mathbfit}{cmr}{bx}{it} 
  \addtoversion{bold}{\mathbfit}{cmr}{bx}{it} 
  \newmathalphabet{\mathbfss} 
\textbfss{..} 
  \addtoversion{normal}{\mathbfss}{cmss}{bx}{n} 
  \addtoversion{bold}{\mathbfss}{cmss}{bx}{n} 
  \ifAMStwofonts 
    \ifCUPmtlplainloaded \else 
and 
your 
      \UseAMStwoboldmath 
      \makeatletter 
      \new@mathgroup\upmath@group 
      \define@mathgroup\mv@normal\upmath@group{eur}{m}{n} 
      \define@mathgroup\mv@bold\upmath@group{eur}{b}{n} 
      \edef\UPM{\hexnumber\upmath@group} 
      \new@mathgroup\amsa@group 
      \define@mathgroup\mv@normal\amsa@group{msa}{m}{n} 
      \define@mathgroup\mv@bold\amsa@group{msa}{m}{n} 
      \edef\AMSa{\hexnumber\amsa@group} 
      \makeatother 
      \mathchardef\upi="0\UPM19 
      \mathchardef\umu="0\UPM16 
      \mathchardef\upartial="0\UPM40 
      \mathchardef\leqslant="3\AMSa36 
      \mathchardef\geqslant="3\AMSa3E 
      \let\oldle=\le     \let\oldleq=\leq 
      \let\oldge=\ge     \let\oldgeq=\geq 
      \let\leq=\leqslant \let\le=\leqslant 
      \let\geq=\geqslant \let\ge=\geqslant 
    \fi 
  \fi 
\fi 
 
\ifnfsstwo 
  \newcommand{\rmn}[1] {\mathrm{#1}} 
  \newcommand{\itl}[1] {\mathit{#1}} 
  \newcommand{\bld}[1] {\mathbf{#1}} 
  \def\textbfit{\protect\txtbfit} 
  \def\textbfss{\protect\txtbfss} 
  
\long\def\txtbfit#1{{\fontfamily{cmr}\fontseries{bx}\fontshape{it}
    \selectfont #1}} 
  
\long\def\txtbfss#1{{\fontfamily{cmss}\fontseries{bx}\fontshape{n
}%
    \selectfont #1}} 
  \DeclareMathAlphabet{\mathbfit}{OT1}{cmr}{bx}{it} 
  \SetMathAlphabet\mathbfit{bold}{OT1}{cmr}{bx}{it} 
  \DeclareMathAlphabet{\mathbfss}{OT1}{cmss}{bx}{n} 
  \SetMathAlphabet\mathbfss{bold}{OT1}{cmss}{bx}{n} 
  \ifAMStwofonts 
    \ifCUPmtlplainloaded \else 
      \DeclareSymbolFont{UPM}{U}{eur}{m}{n} 
      \SetSymbolFont{UPM}{bold}{U}{eur}{b}{n} 
      \DeclareSymbolFont{AMSa}{U}{msa}{m}{n} 
      \DeclareMathSymbol{\upi}{0}{UPM}{"19} 
      \DeclareMathSymbol{\umu}{0}{UPM}{"16} 
      \DeclareMathSymbol{\upartial}{0}{UPM}{"40} 
      \DeclareMathSymbol{\leqslant}{3}{AMSa}{"36} 
      \DeclareMathSymbol{\geqslant}{3}{AMSa}{"3E} 
      \let\oldle=\le     \let\oldleq=\leq 
      \let\oldge=\ge     \let\oldgeq=\geq 
      \let\leq=\leqslant \let\le=\leqslant 
      \let\geq=\geqslant \let\ge=\geqslant 
    \fi 
  \fi 
\fi 
 
\ifCUPmtlplainloaded \else 
  \ifAMStwofonts \else 
    \def\upi{\pi} 
    \def\umu{\mu} 
    \def\upartial{\partial} 
  \fi 
\fi 
 
   \title[Direct Calibration of the Cepheid Period-Luminosity relation]{ 
Direct Calibration of the Cepheid Period-Luminosity relation
\thanks{Based on data from the ESA HIPPARCOS astrometry satellite}
}
 
   \author[Lanoix P. et al.] 
{Lanoix P. $^{1,2}$, Paturel G. $^1$, Garnier R.$^1$\\
$^1$ CRAL-Observatoire de Lyon,
     F69230 Saint-Genis Laval, FRANCE,\\
$^2$ Universit\'e Claude Bernard Lyon I, 
     F69622 Villeurbanne, FRANCE\\}

   \date{Received December 1998; accepted -- -- --}
\pubyear{1998}
 
\begin{document}
\maketitle
 
\begin{abstract}
After the first release of HIPPARCOS data, Feast \& Catchpole gave a new
value to the zero-point of the visual Cepheid Period-Luminosity relation based 
on trigonometric parallaxes. Because of the large uncertainties on these 
parallaxes, the way in which individual measurements are weighted bears
a crucial importance, 
and the discrepancy they show leads to the conclusion that the choice of
the best weighting system can be provided through a Monte-Carlo simulation. \\
On the basis of such a simulation it is shown that: 
\begin{itemize}
\item A cut in $\pi$ or in $\sigma_{\pi} / \pi$ introduces a strong bias.
\item The zero-point is more stable when only the brightest Cepheids are used.
\item The Feast \& Catchpole weighting gives the best zero-point and the lowest dispersion.
\end{itemize}
After correction, the adopted visual Period-Luminosity relation is:
$$ \langle M_V \rangle = -2.77 \log P - 1.44 \pm 0.05.$$
Moreover, we extend this study to the photometric I-band (Cousins) and obtain:
$$ \langle M_I \rangle = -3.05 \log P - 1.81 \pm 0.09.$$ 
\end{abstract}


\section{Introduction}

Cepheid variables constitute one of the
most important primary distance calibrators. 
Indeed, they obey a Period-Luminosity (PL) relation:
\begin{equation}
\langle M_V \rangle = \delta \  \log P + \rho
\end{equation}
 from which the absolute magnitude $\langle M_V \rangle$
can be determined just from the measurement of the period,
provided that the slope $\delta$ and the zero-point $\rho$ are known.

The slope of the PL relation seems very well established 
from ground-based observations in the Large Magellanic Cloud (LMC) because
the population incompleteness bias pointed out for more distant galaxies 
(Lanoix et al. 1999a) seems negligible in the LMC. 
The slope of the PL relation is easier to obtain from an external galaxy
because, all Cepheids being at the same distance, the slope can be
determined by using apparent magnitudes instead of absolute magnitudes.
A reasonable value for the photometric V-band is 
$\delta = -2.77 \pm 0.08$ (see for instance Gieren et al. 1998, Tanvir 1997, Caldwell \& Laney 1991, Madore 
\& Freedman 1991).
In the present study we will adopt this value and will discuss further the effect
of a change of it.

The establishment of the zero-point still remains a
major goal.  Today, thanks to the HIPPARCOS satellite
\footnote{HIPPARCOS parallaxes are ten times better 
than those obtained from ground-based observations
(i.e., $\sigma_{\pi} \approx 1$ milliarcsec).}, 
the trigonometric parallaxes of galactic Cepheids are accessible, 
allowing a new determination of $\rho$.

After the first release of HIPPARCOS data, a 
calibration of the Cepheid PL relation 
was published by Feast \& Catchpole (1997, hereafter FC).
This work gave a distance for the LMC galaxy larger than
the one generally assumed. 
However, some papers (Madore \& Freedman 1998, Sandage \& Tammann 1998)
argued that this calibration is only brighter than previous ones
at the level of $\le 0.1$ mag.
An independent study of the calibration of the PL relation
based on the same data also led to a long distance scale (Paturel et al. 1996)
and to a large LMC distance modulus of 18.7 (Paturel et al. 1997).
All these studies may be affected by statistical biases due either
to the cut of negative parallaxes or to the method used for bypassing
these cuts. This justifies that we want to analyze deeper these
results.

HIPPARCOS parallaxes $\pi$ may have large standard deviations $\sigma_{\pi}$
leading sometimes to negative parallax so that the distance $d$(pc)=1/$\pi$ cannot 
be calculated. Anyway, it is a biased estimate of the true distance 
(Brown et al. 1997). Thus, it seems impossible to use it for a direct
calculation of the zero-point. 
On the other hand, rejecting negative parallaxes 
generates a Lutz-Kelker bias type
(Lutz \& Kelker 1973) while rejecting parallaxes with
large $\sigma_{\pi}/{\pi}$ 
generates another bias (Brown et al. 1997). 
In order to bypass this problem, FC suggest
calculating $\rho$ from the weighted mean of the function:
\begin{equation}
10^{0.2\rho} = 0.01 \pi 10^{0.2(\langle V_0 \rangle - \delta \log P)}
\label{rho}
\end{equation}
This treatment assumed that the exponent of a mean is identical to the
mean of the exponents. FC justify it by saying that ``the scatter about
the PL(V) relation is relatively small''. 
They chose a weighting and compute the mean 
of $10^{0.2 \rho}$, from which they derive $\rho$. \\
As a matter of fact, they use a Period-Color (PC) relation for dereddening their magnitudes.
Because of the near degeneracy of the reddening slope and the colour term in a 
Period-Luminosity-Color relation, this technique will have much the same narrowing effect
on the PL relation as including a color term would. For a Cepheid of known distance 
the scatter is reduced from 0.2 down to about 0.1. 
However since the HIPPARCOS parallaxes may have large errors, we see from equation 
\ref{rho} that the scatter in $10^{0.2\rho}$ could be increased in this manner.
      
Precisely, we would like to answer the following questions:
\begin{itemize}
\item Can we obtain a good result by rejecting poor parallaxes?
\item Is the dispersion small enough to justify the calculation of $\rho$ using the
mean of $10^{0.2 \rho}$?
\item Is the final result biased or not?
\item Is it possible to adopt another weighting than that of FC?
\end{itemize}

In section 2 we use the HIPPARCOS sample of Cepheids to confirm that
rejecting negative parallaxes or parallaxes with a poor $\sigma_{\pi} / \pi$
gives a biased zero-point and to test the FC method with different 
weighting systems.  
This suggests making a simulated sample for 
which the zero-point is {\it a priori} known and then to apply the same 
treatment to it. 

In section 3 we explain how the simulated sample is built in order to
reproduce all the properties of the true HIPPARCOS sample.

Then, in section 4 we give the result of the FC method applied to the simulated 
sample with different weightings. This shows that the calculated zero-points 
and the associated standart deviations depend on 
the adopted weigthing.

In section 5, the previous results are discussed
and explained. The consequences are drawn for estimating
the best zero-point from the HIPPARCOS Cepheid sample for both V and I bands.

\section{Use of the HIPPARCOS Cepheid sample}
The complete Cepheid sample  is extracted from the catalogue 
HIPPARCOS (1997). Among all variable stars, we keep only those labelled DCEP 
(classical $\delta-$type Cepheids) and DCEPS (first overtone pulsators), 
and then obtained a total of 247 Cepheids. The period of the 31 overtone pulsators
is converted to the fundamental period $P$ according to Alcock et al. (1995):
\begin{equation}
P_{1}/P=0.716 - 0.027 \log P_{1}
\end{equation}

The $B$ and $V$ photometry is available from the David Dunlap Observatory Galactic Cepheid
Database (Fernie et al. 1995), except for nine Cepheids 
(CK Cam, BB Gem, KZ Pup, W Car, DP Vel, BB Her, V733 Aql, KL Aql and V411 Lac)
which were excluded from the present study.
Therefore, the final sample (table 4) is made of 238 Cepheids (31 overtones).

\begin{figure}
\epsfxsize=8.5cm
\epsfbox{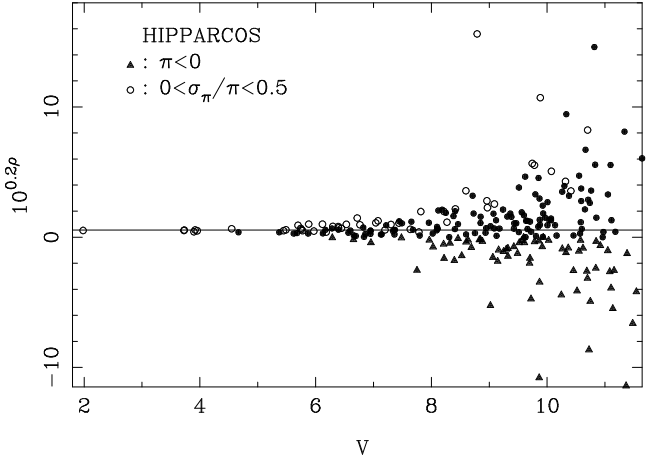}
\caption{Effects of cuts. The horizontal line corresponds to the
zero-point value $\rho = -1.43$. 
If one rejects negative parallaxes (filled triangles) or keeps parallaxes with
$0 \le \sigma_{\pi} /  \pi \le 0.5$ (open circles), the mean is overestimated.
}
\label{cut}
\end{figure}

The color excess is then calculated using the FC method,
i.e. calculation of the intrinsic color $\langle B \rangle _0- \langle V
\rangle _0$ from a linear relation
color vs. $\log P$, according to Laney \& Stobie (1994):
\begin{equation}
\langle B \rangle _0- \langle V \rangle _0 = 0.416 \log P +0.314.
\label{laney}
\end{equation}
We use the relation from Laney \& Stobie (1993) to compute the V extinction :
\begin{equation}
R_{V} = 3.07 + 0.28 (\langle B \rangle _0- \langle V \rangle _0) + 0.04 E_{(B-V)}
\label{laney2}
\end{equation} 
Figure \ref{cut} shows how the quantity $10^{0.2 \rho}$ varies with the apparent 
magnitude $V$. This quantity is directly
needed for the calculation of the zero-point $\rho$. Clearly, the dispersion
increases with the magnitude, but the distribution is quite symetrical
around a given value. 

If a cut is applied on the sample to reject negative parallaxes
(filled triangles in figure \ref{cut}) the mean of $10^{0.2 \rho}$ is 
overestimated. If one uses only measurements with 
$0 < \sigma_{\pi} / \pi < 0.5 $
(open circles in figure \ref{cut}), again, $10^{0.2 \rho}$ is  overestimated.  
Thus, as claimed by Brown et al. (1997), a bias is clearly 
confirmed if one cuts the sample. We will no more consider cuts involving parallaxes
as a way of obtaining a valuable result.

Figure \ref{cut} does not exhibit 
a small dispersion. So, we do not know if the FC's
procedure leads to the proper value of $\rho$.
For the calculation of the mean of $10^{0.2 \rho}$ they
use individual weights taken as the reciprocal of the square of the
standard error of the second term of equation \ref{rho}.
For a given Cepheid, the weight is given by: 
\begin{equation}
\omega_{i} \approx  [10^{-2} \sigma_{\pi_{i}} 10^{0.2(\langle V_{0_{i}} \rangle - \delta \log P_{i})}]^{-2}
\label{wfc}
\end{equation}
because the error on the term $10^{0.2(\langle V_{0_{i}} \rangle 
- \delta \log P_{i})}$ is negligible as shown by FC.
This weighting is mathematically the most rigorous. However, some
other empirical weightings may be worthy of interest. 

Since the error on $\rho$ is mainly due to the large uncertainty $\sigma_{\pi}$, 
we will test a weight in $\sigma_{\pi_{i}}^{-2}$ and in $(\sigma_{\pi_{i}}/\pi_i)^{-2}$.
Further, we will also use an unweighted mean because the dispersion looks quite
symetrical around a mean value and a $V^{-2}$ weighting because the dispersion increases
with V.
We then repeated the FC tests as well as the other weightings and found the results
given in table \ref{weight}. 

\begin{table}
\caption{Values of $\rho$ calculated with different weightings and different
cuts in $V$ magnitude. The standard deviation of each value is given in
parenthesis.}
\begin{tabular}{lrrr}
\hline
Weighting                        &         All $V$  &    $V \leq 8$   &  $V \leq 6$      \\
\hline
F\&C                             &  $ -1.45(0.10)$  &  $ -1.47(0.10)$ &  $ -1.45(0.08)$  \\
$\sigma_{\pi_{i}}^{-2}$          &  $ -1.04(0.37)$  &  $ -1.38(0.22)$ &  $ -1.45(0.16)$  \\
$(\sigma_{\pi_{i}}/\pi_i)^{-2}$  &  $  1.19(0.57)$  &  $      -     $ &  $      -     $  \\
No weight                        &  $ -0.19(0.74)$  &  $ -1.38(0.22)$ &  $ -1.40(0.17)$  \\
$V^{-2}$                         &  $ -0.64(0.65)$  &  $ -1.41(0.22)$ &  $ -1.42(0.13)$  \\
\hline
\label{weight}
\end{tabular}
\end{table}

From this table we see that, when all Cepheids are used, the calculated 
zero-point $\rho$ strongly depends on the adopted weighting.
The instability of this result can be explained by the very large dispersion
at large $V$. This large dispersion quite justifies the second question of
section 1.

According to the shape of figure \ref{cut}, we see that the dispersion can be 
reduced by cutting the sample at a given apparent magnitude. Table \ref{weight} shows that such
a cut gives a more stable result.
Moreover, the weighting adopted by FC gives the lowest dispersion. 
For instance, keeping the brightest 11 Cepheids, we obtain $\rho=-1.45$ 
with a very small standard deviation of 0.05 ($V \leq 5.5$). We also try to
keep only stars with the highest weights (whatever the weighting system). However, that leads us to
the same results with slightly higher dispersions. 

In practice, we have no means of knowing if a bias has been introduced
as long as the observed sample is used because the true zero-point is
not known. Only a simulated sample, with a zero-point {\it a priori} known,
can provide the answer to the third question of section 1.
This justifies the construction of simulated samples.

\section{Construction of simulated samples}
To build a simulated sample only three quantities have to be drawn independently:
\begin{itemize}
\item The parallax $\pi$
\item The logarithm of the period $\log P$
\item The column density of interstellar matter along the line of sight.
\end{itemize}

\subsection{The simulated ``true parameters''}
\begin{figure}
\epsfxsize=6.5cm
\epsfbox{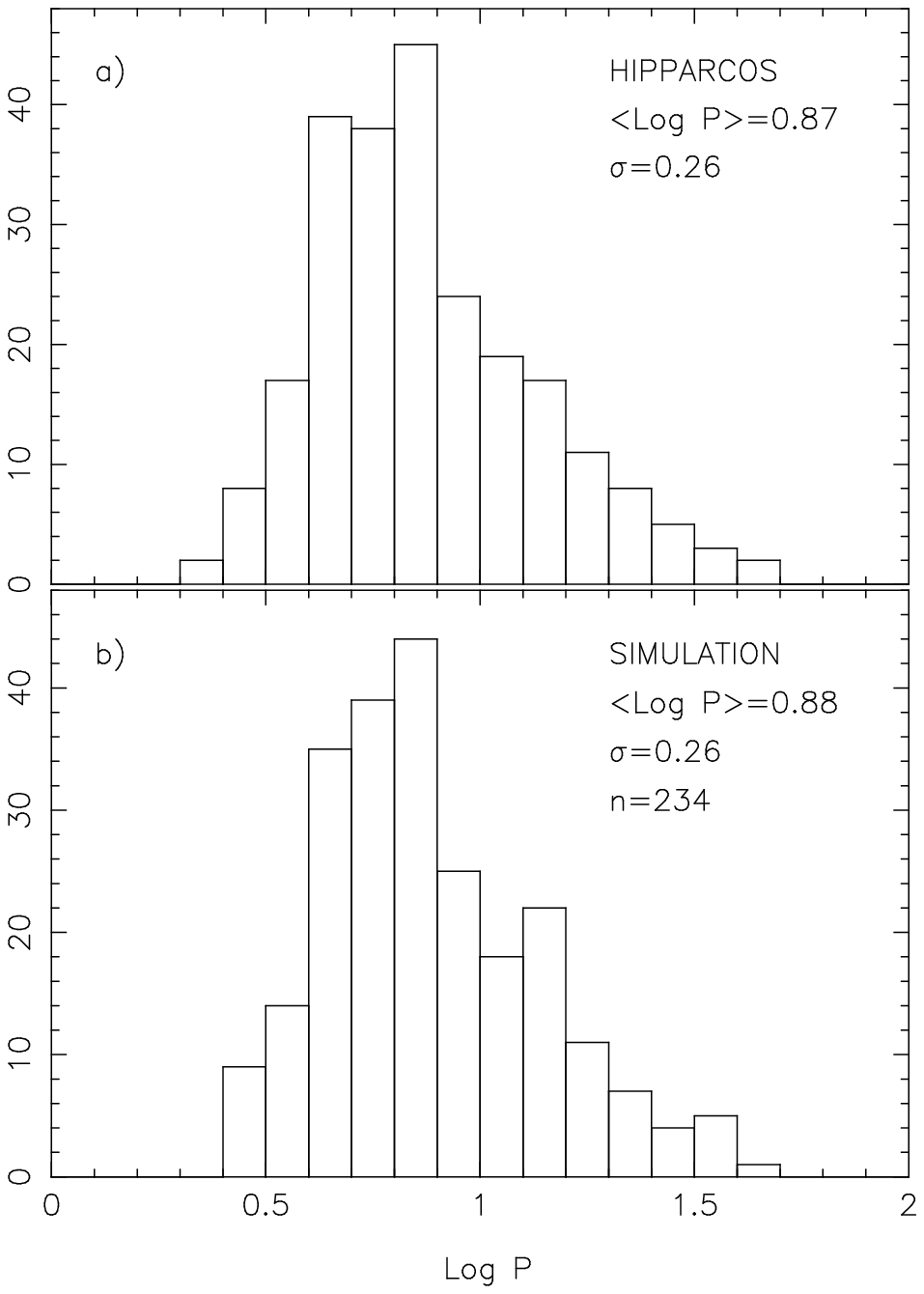}
\caption{Histograms of $\log P$ for HIPPARCOS data and for a simulated sample.
In Figure a, the periods of overtone pulsators are corrected (see text). 
}
\label{h_logp}
\end{figure}

Assuming a homogeneous 3D distribution of galactic 
Cepheids (this is justified owing to 
relatively small depth of HIPPARCOS survey regarding the 
depth of the galactic disk),
we draw at random the x,y,z coordinates over the range 
[-2100, 2100] pc. We keep only
Cepheids within a radius of 2100 pc and then deduce the true parallax:
\begin{equation}
\pi = 1/ \sqrt{x^2+y^2+z^2}
\end{equation}
250 true parallaxes are drawn in such a way. Each point will 
be a Cepheid in our simulated sample.\\
Then, for each Cepheid  we draw $\log P$ following a distribution which reproduces 
the observed distribution of periods (Fig. \ref{h_logp}a and  \ref{h_logp}b). 
We then calculate the absolute magnitude $\langle M_V \rangle$ 
from the relation:
\begin{equation}
\langle M_V \rangle = \delta_V \log P + \rho_V + \Delta
\end{equation}
where $\delta_V = -2.77$ is the adopted slope as said in the introduction,
$\rho_V= -1.30 $ is the 
arbitrarily fixed zero-point and $\Delta$ is a Gaussian
intrinsic dispersion ($\langle \Delta \rangle =0$ ; $\sigma(\Delta)=0.2$) 
which reflects the width of the instability strip.
The absolute magnitude in B-band $\langle M_B \rangle$ is calculated 
in the same way using
the same intrinsic dispersion multiplied by 1.4.
We reproduce in this manner the correlation of the residuals
as well as the dispersion of the true CP relation related to
the color variation across the instability strip. We chose 
$\delta_B=\delta_V+0.416$ and $\rho_B=\rho_V + 0.314$,
so that it implies the relation between the intrinsic 
color $\langle B \rangle _0 - \langle V \rangle _0 $ and
$\log P$ from Laney \& Stobie (1994):
\begin{equation}
\langle B \rangle _0 - \langle V \rangle _0 = (\delta_B - \delta_V) \log P
+ \rho_B - \rho_V
\label{color}
\end{equation}
The true intrinsic color $\langle B \rangle _0 - \langle V \rangle _0$
 is calculated from this linear relation. We then reduce the dispersion
of the PL relation down to 0.1 as already explained in the introduction.

\begin{figure}
\epsfxsize=8.5cm
\hbox{\epsfbox{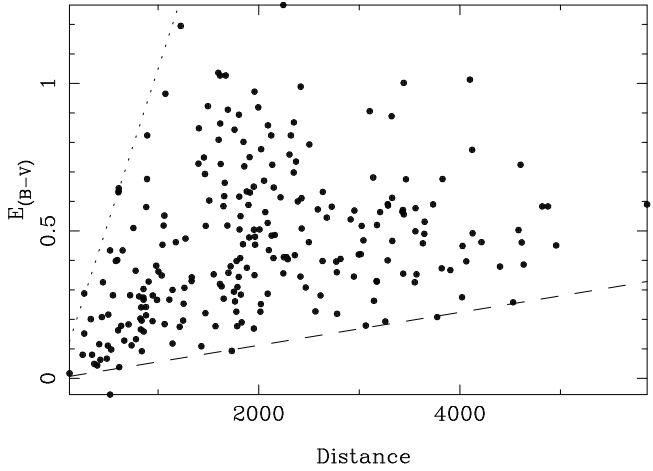}}
\caption{Color exces versus photometric distances from Fernie (Fernie et al.
1995) for HIPPARCOS Cepheids. 
The slope $E_{(B-V)} / distance $ measures the density of the interstellar
medium. In slightly obscured directions (dashed line) 
one can see stars up to $\approx 5000$ pc, while in very obscured regions 
(dotted line) the closest Cepheids are detected not farther than $\approx 1100$ pc. 
}
\label{ebmv_dist}
\end{figure}

The relation of $E_{(B-V)}$ versus the calculated photometric 
distances (adopting, for instance, distances from Fernie et al. 1995) shows (Fig. \ref{ebmv_dist}) 
that the observed Cepheids are located in
a sector. All line of sight directions have extinction (no point below
the dashed line). In slightly obscured directions (dashed line) one can see
stars up to $\approx 5000$ pc, while in very obscured regions (dotted line) 
the closest Cepheids are detected not farther than $\approx 1100$ pc. 

The slope $E_{(B-V)} / distance $ is a 
measure of the density of the interstellar
medium in a given direction. This density varies over a large range
due to the patchiness of the galactic extinction, but, for a given
line of sight, the extinction, and thus the color excess, is assumed to be
proportional to the distance. This figure is used to obtain the extinction
for each Cepheid.  We draw at random the slope $E_{(B-V)} / distance $ over 
the range defined by the dashed and dotted lines (Fig. \ref{ebmv_dist}). 
Using the true distance $1/\pi$ we then deduce the true color excess 
$E_{(B-V)}$, and the true extinctions:
\begin{equation}
A_V= R_V E_{(B-V)}
\end{equation}
\begin{equation}
A_B= R_B  E_{(B-V)}
\end{equation}
 with $R_V$ = 3.3 and $R_B = 4.3$.

\subsection{The simulated ``observed parameters''}
Now we calculate the parameters which would be observed.
First, the apparent $B$ and $V$ magnitudes
are simply:
\begin{equation}
\langle V \rangle = 5 \log (1/\pi) - 5 + \langle M_V \rangle + A_V + \epsilon_V
\end{equation}
\begin{equation}
\langle B \rangle = 5 \log (1/\pi) - 5 + \langle M_B \rangle + A_B + \epsilon_B
\end{equation}
where $\epsilon_V$ and $\epsilon_B$ are two independent Gaussian variables
which reproduce measurement uncertainties (the intrinsic 
scatter of the PL relation
is already counted in $\langle M_V \rangle $ and $\langle M_B \rangle $). We adopted for both:
$\langle \epsilon \rangle = 0.0$ and $\sigma_{\epsilon} = 0.005$.

\begin{figure}
\epsfxsize=6.5cm
\hbox{\epsfbox{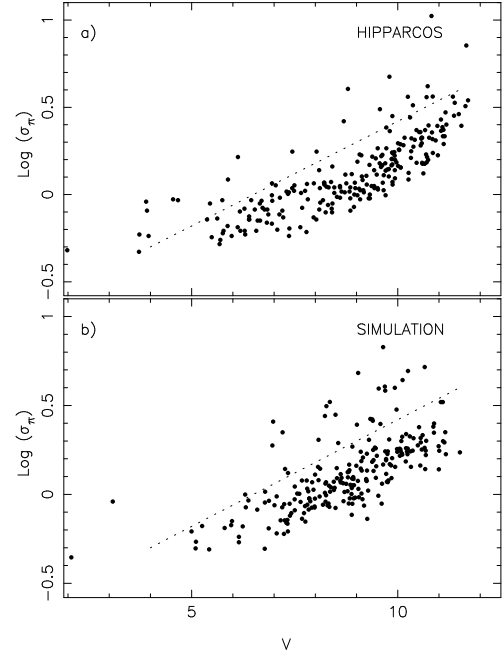}}
\caption{$\log \sigma_{\pi}$ vs. $V$ for HIPPARCOS data and for a simulated sample.
We can see the two populations of Cepheids as described in the text.
}
\label{spi_v}
\end{figure}

The parallax which would be observed is calculated from the true one
and an associated $\sigma_{\pi}$ obtained through
the figure \ref{spi_v}a. This figure shows two populations: one below the
dotted line, the other about the dotted line.
First, we draw the membership to one of these families in the right proportion.
Then, from the linear relationships of the corresponding family 
and the $V$ magnitude already computed, we calculate
$\log \sigma_{\pi}$ (i.e. $\sigma_{\pi}$). Finally, the 
observed $\pi$ is obtained
by drawing one occurence in the Gaussian distribution $(\pi, \sigma_{\pi})$.\\
Concerning the observed color excess, it will simply 
be deduced from the relation:
\begin{equation}
E_{(B-V)}= \langle B \rangle - \langle V \rangle - 
 (\langle B \rangle _0 - \langle V \rangle _0) 
\label{exces}
\end{equation}
with $\langle B \rangle _0 - \langle V \rangle _0$ deduced from the PC 
relation \ref{color} as we did in section 2.

We also need to determine the observed value of the coefficient $R_V$.
We draw its value according to a Gaussian distribution centered on the chosen
true value (3.3) with a dispersion of 0.05. So, we suppose that the observed
value has no systematic shift with respect to the true value.

Finally, in order to reproduce selection effects like the Malmquist bias (Malmquist 1920)
we reject the Cepheids which could not be observed according to their
apparent magnitudes (i.e. their probability to be detected). We draw a
random parameter $t \in [0, 1]$ and compute the quantity:
\begin{equation}
t_0 = \frac{1}{1+\exp^{\alpha (\langle V \rangle - \langle V_{lim} \rangle)}} 
\end{equation}
Whenever $t \le t_0$ the star may be observed by HIPPARCOS and we keep 
it in our sample, and in the other case it will be rejected. We assume 
$\alpha = 1$ and $\langle V_{lim} \rangle = 12.5$. Moreover, whenever  
$ \langle V \rangle \le 1.9$, the Cepheid would be too bright
(unrealistic apparent magnitude) and then rejected. The number of
simulated Cepheids is then almost equal to the true one.

\begin{figure}
\epsfxsize=6.5cm
\hbox{\epsfbox{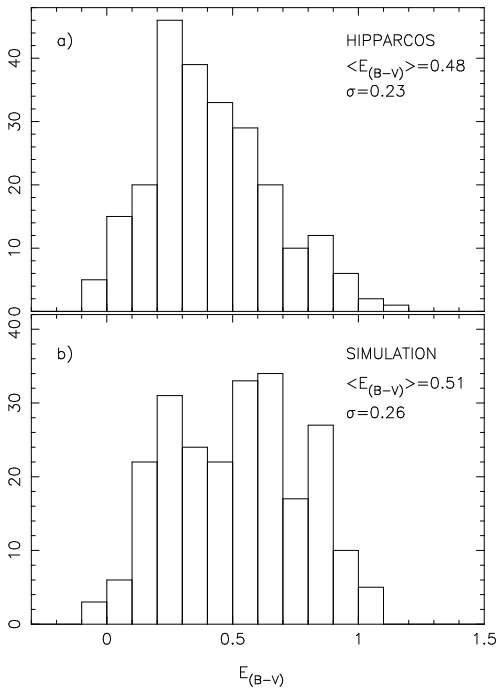}}
\caption{Histograms of observed $E_{(B-V)}$ for HIPPARCOS
 data and for a simulated sample
according to equations \ref{laney} and \ref{exces} respectively.
}
\label{h_ebmv}
\end{figure}

\begin{figure}
\epsfxsize=6.5cm
\hbox{\epsfbox{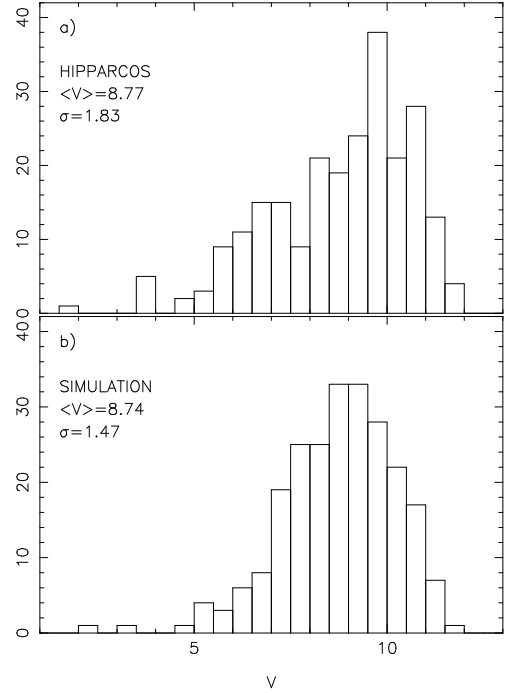}}
\caption{Apparent $V$ magnitudes for HIPPARCOS data and for a simulated sample.
}
\label{h_v}
\end{figure}

\begin{figure}
\epsfxsize=6.5cm
\hbox{\epsfbox{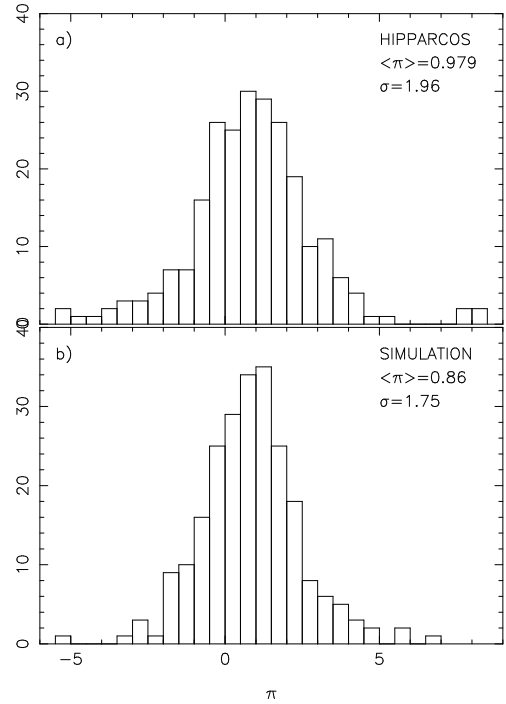}}
\caption{Observed parallaxes $\pi$ in mas for HIPPARCOS data and 
for a simulated sample.
}
\label{h_pi}
\end{figure}

\begin{figure}
\epsfxsize=6.5cm
\hbox{\epsfbox{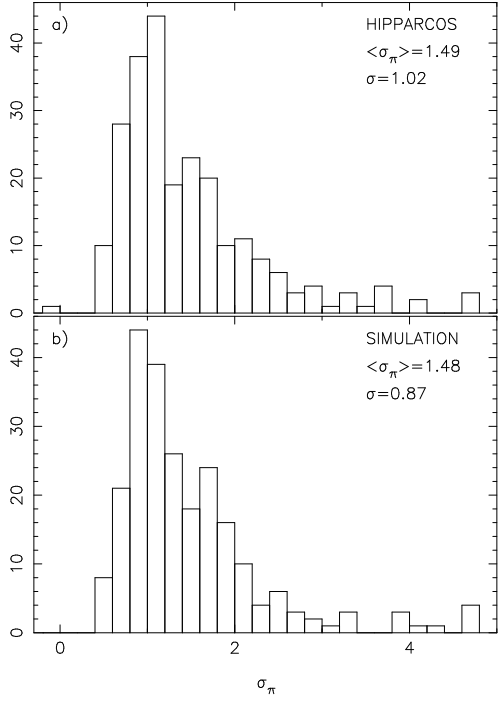}}
\caption{Errors on the observed parallaxes for HIPPARCOS data and 
for a simulated sample.
}
\label{h_spi}
\end{figure}

\begin{figure}
\epsfxsize=6.5cm
\hbox{\epsfbox{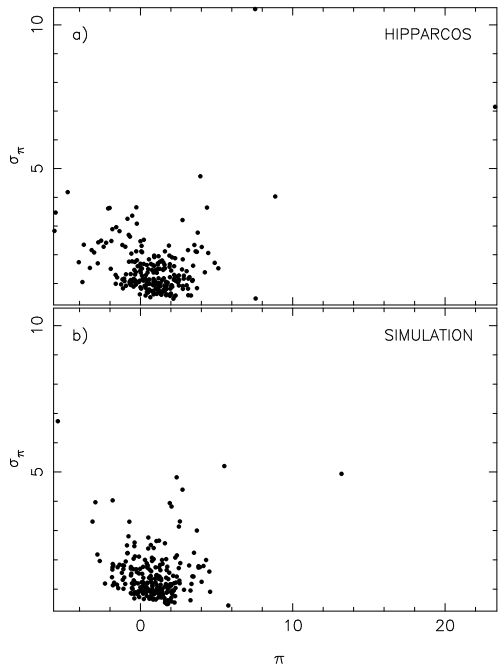}}
\caption{$\sigma_{\pi}$ vs. $\pi$ for HIPPARCOS data and for a simulated sample.
The two quantities are not correlated.
}
\label{spi_pi}
\end{figure}

\begin{figure}
\epsfxsize=6.5cm
\hbox{\epsfbox{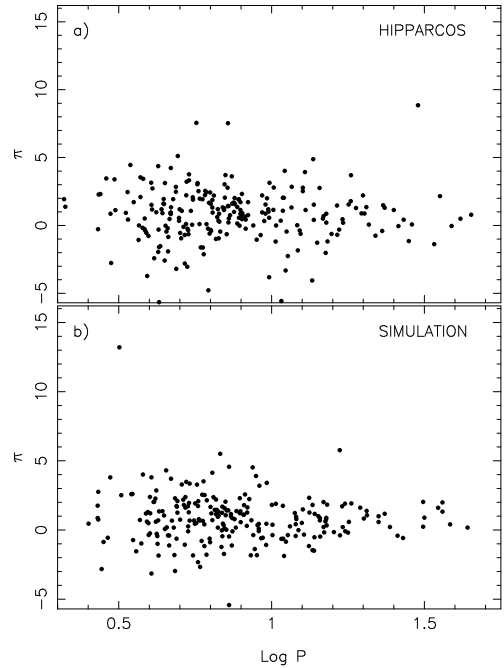}}
\caption{$\pi$ vs. $\log P$ for HIPPARCOS data and for a simulated sample. It 
shows that there's no correlation between these two quantities. 
}
\label{pi_logp}
\end{figure}

\begin{figure}
\epsfxsize=6.5cm
\hbox{\epsfbox{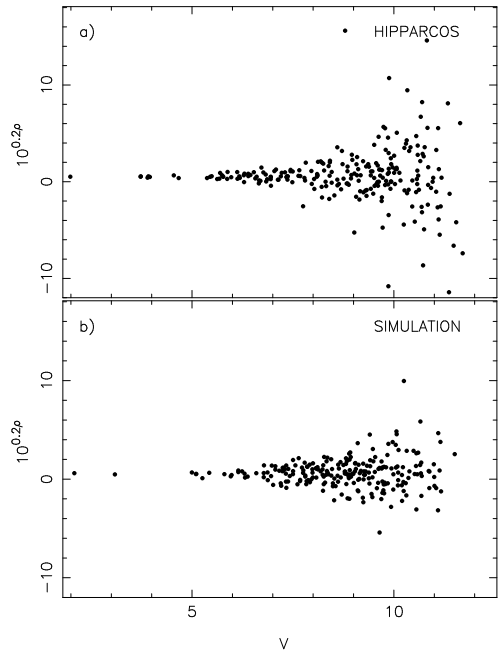}}
\caption{The distribution
of the exposant of the zero-points is plotted as a function of the apparent
$V$-magnitude for both HIPPARCOS data and a simulated sample. 
Note that figure a is the same than figure \ref{cut}.
}
\label{rho_v}
\end{figure}

In order to show that the simulated sample is comparable to the
true HIPPARCOS one, we plot for one simulated sample the 
same figures (Fig. \ref{h_logp} to \ref{rho_v}) as those produced 
with the true HIPPARCOS sample. Note that the figures from the
simulated sample are made from a single drawing which is not necessarily
an optimal representation of the true sample.

\section{RESULTS}
The result may depend on the particular sample we draw. In order to reduce the
uncertainty due to this choice, we made 1000 different random drawings 
(each of them with about 240 Cepheids) and adopted the mean result. 
We obtain the result shown in the table 2 (let us recall that the input zero-point
is $\rho_{V}=-1.30$).

\begin{table}
\caption{Values of $\rho$ calculated using 1000 simulated samples. We used different 
weightings and different cuts in $V$ magnitude as in the study made with the true sample. 
The standard deviation of each value is given in parenthesis.}
\begin{tabular}{crrr}
\hline
weighting                        &         All $V$  &    $V \leq 8$   &  $V \leq 6$      \\
\hline
true zero-point                  &  $ -1.30 $       &  $ -1.30 $      &  $ -1.30 $       \\
\hline
F\&C                             &  $ -1.31(0.14)$  &  $ -1.30(0.15)$ &  $ -1.31(0.21)$  \\
$\sigma_{\pi_{i}}^{-2}$          &  $ -1.33(0.21)$  &  $ -1.31(0.22)$ &  $ -1.31(0.26)$  \\
$(\sigma_{\pi_{i}}/\pi_i)^{-2}$  &  $  0.03(0.47)$  &  $      -     $ &  $      -     $  \\
No weight                        &  $ -1.36(0.43)$  &  $ -1.33(0.39)$ &  $ -1.32(0.34)$  \\
$V^{-2}$                         &  $ -1.33(0.33)$  &  $ -1.32(0.32)$ &  $ -1.31(0.29)$  \\
\hline
\end{tabular}
\end{table}

The simulation clearly confirms that the weighting in 
$(\sigma_{\pi_{i}}/\pi_i)^{-2}$ is meaningless. 
Again, it is confirmed that a cut in magnitude gives more stable results
because the method of averaging $10^{0.2 \rho}$ to get $\rho$ is better
justified with small dispersion. 
This answers the second question of section 1.
The simulation also confirms that the FC weighting leads to the lowest dispersion 
and that the results are too low at only a 0.02 or 0.01 mag. level.

\begin{figure}
\epsfxsize=8.5cm
\hbox{\epsfbox{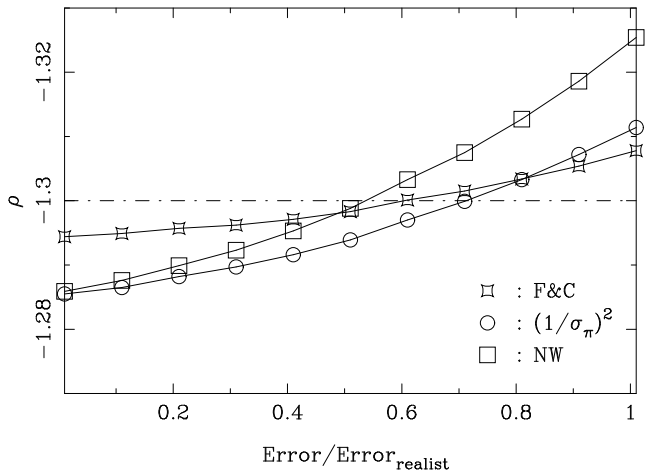}}
\caption{Zero-point values for each weighting when the observational error
is progressively reduced from its realist value down to zero.}
\label{rho_err}
\end{figure}

In order to analyze the effect of the measurement errors,
we progressively reduce the observational 
errors (but not the intrinsic dispersion) introduced in our simulation. The reduction
is made from their realistic values down to zero. We compute the mean value of the
distribution of $\rho$ as we go along, and plot the results in figure~\ref{rho_err}.  
It appears that the zero-point values comes closer to the
real value $\rho = -1.30$. Moreover, the FC weighting gives clearly the
more stable result.
The trends of figure \ref{rho_err} (decreasing of $\rho$ with increasing errors)
can be explained solely by errors on $\pi_{i}$ because they disappear when
$\sigma_{\pi_{i}}$ is forced to zero.

Further, we checked that removing both the measurement errors and the intrinsic dispersion
removes the residual shift for all kinds of weighting and gives back the initial
value $\rho = -1.30$.
This proves that our simulation procedure works well.

\section{DISCUSSION}
The results of the previous section allow us to answer the
questions of section 1:
a cut in apparent magnitude reduces the dispersion and
gives reliable results because averaging $10^{0.2 \rho}$ works 
better with small dispersion.
Whatever the weighting adopted, the zero-point is not biased by more than $0.03$ mag.
The FC weighting gives the
smallest standard deviation, and the systematic shift never exceeds $0.01$ mag.

Let us analyze the main effects which are responsible for a shift. Two effects are 
present: effect of averaging in $10^{0.2 \rho}$ and Malmquist effect. We will see
that they work in two opposite directions.

Consider two Cepheids comparable in every aspect, 
i.e. located at the same distance in two
directions with the same interstellar absorption, measured with the same
$\sigma_{\pi}$ so that they have the same observed parallaxes, 
and both with the same periods, but one located near one edge of the 
instability strip whereas the second is located at the opposite edge.
Their absolute magnitudes would then be:
\begin{equation}
\langle M_V^1 \rangle = \delta \log P + \rho + \zeta
\end{equation}
\begin{equation}
\langle M_V^2 \rangle = \delta \log P + \rho - \zeta
\end{equation}
where $\zeta$ is the actual value of the intrinsic dispersion 
($\langle \Delta \rangle =0$ ; $\sigma(\Delta) \approx 0.2$) across 
the instability strip.  \\
When using these two Cepheids to compute the zero-point of the PL relation
directly from the parallax we would obtain:
\begin{equation}
\rho_1 = \langle V_{0} \rangle + \zeta' + 5 \log \pi -10 - \delta \log P 
\label{rho1}
\end{equation}
\begin{equation}
\rho_2 = \langle V_{0} \rangle - \zeta' + 5 \log \pi -10 - \delta \log P
\label{rho2}
\end{equation}
with $\zeta' \le \zeta$ because of dereddening method.
The mean of the two values corresponds then to the true value $\rho$ since:
\begin{equation}
\frac{\rho_1 + \rho_2}{2} = \rho_{true}
\end{equation}
However we have shown why such a direct mean cannot be used with HIPPARCOS data.
So we compute the mean (or the weighted mean) of the two quantities 
$10^{0.2 \rho_1}$ and $10^{0.2 \rho_2}$. 
The mean zero-point $\overline{\rho}$ can be expressed as:
\begin{equation}
\overline{\rho} = \rho_{true} + 
5 \ \log [\frac{{\omega_1}/{\omega_2}10^{0.2 \zeta'} + 10^{- 0.2 \zeta'}}
{1 + {\omega_1}/{\omega_2}} ]
\end{equation}
where $\omega_1$ and $\omega_2$  are the weights of the two quantities.

If we adopt $\omega_1/\omega_2=1$ (i.e. no weighting or same weights) and $\zeta'=0.2$
(overvalued in order to highlight the way $\rho$ is biased) 
we obtain $\overline{\rho} = \rho_{true} + 0.01$. The observed $\rho$ slighly increases 
in this manner.

The Malmquist effect works in the other direction. The biased absolute magnitude $\overline{M'}$
is too bright (Teerikorpi 1984) :
\begin{equation}
\overline{\langle M' \rangle} = \overline{\langle M \rangle} - 1.38 \sigma^{2}
\end{equation}
where $\overline{\langle M \rangle}$ is the unbiased magnitude. This formula gives the global 
correction, not the correction for individual Cepheid. Assuming a pessimistic value $ \sigma = 0.2$
(once again, since the use of the PC relation as a narrowing effect, $ \sigma $ is surely lower than
this value) the shift would be at worst $-0.055$. Then the observed $\rho$ diminishes.

Finally, the net shift would be $-0.04$ or less. However, figure \ref{rho_err} which reproduces both
effects with realistic uncertainties gives a shift of $\rho_{observed} = \rho_{true} - 0.01$
when the FC weighting is used. This shift takes in account these two effects. One will then apply
it on the value deduced from HIPPARCOS data.

We investigate now the effect of a change in the adopted slope. We adopted $\delta = -2.77 \pm 0.08$.
What would be the change in the PL relation if the true slope was different from this value?
In table \ref{slope} we give the values of the mean $\langle M_{V} \rangle $ deduced from our simulation, the
input relations being :
\begin{equation} 
\langle M_V \rangle + 4.07 = -2.77 (\log P - 1)
\end{equation}
or
\begin{equation} 
\langle M_V \rangle + 3.74 = -2.77 (\log P - 0.88) 
\end{equation}
We note that the absolute magnitude at $\log P = 1$ (or $\log P = 0.88$) doesn't change very much 
(less than $0.03$) as far as the $\log P$ does not change from the mean of calibrating Cepheids.

\begin{table}
\caption{Effect of the chosen slope on the final magnitudes computed at the mean $\log P$ ($0.88$)
and at $\log P = 1$}
\begin{tabular}{lll}
\hline
slope         &  $\langle M_V \rangle$ at $\log P$ mean & $\langle M_V \rangle$ at 10 d     \\
\hline
-2.60 & -3.76  & -4.07 \\
-2.70 & -3.75  & -4.08 \\
-2.77 & -3.75  & -4.08 \\
-2.80 & -3.75  & -4.08 \\
-2.90 & -3.74  & -4.09 \\
-3.00 & -3.74  & -4.10 \\
\hline
Reference values& & \\
\hline
-2.77 & -3.74 & -4.07 \\
\hline
\label{slope}
\end{tabular}
\end{table}          

\section{CONCLUSION}
The conclusion is that the intrinsic dispersion (even Gaussian and symetrical) of 
the instability strip is responsible for too low values of $\rho$ and may lead 
to a slightly biased result as long as the zero-point $\rho$ is deduced
by averaging $10^{0.2 \rho}$. However it is compensated by the Malmquist bias, and, 
using a PC relation for dereddening the individual Cepheids,
the final effect is globally very small. Indeed, our simulation 
shows that it is almost negligible (Fig. \ref{rho_err}) even when we
account for measurement errors. 
With realistic measurement errors the bias is about $-0.01$. 
A cut in apparent magnitude reduces the uncertainty on the zero-point.
The best unbiased zero-point is obtained by cutting the sample at $V \leq 5.5$ mag.
The result is (after correction of the residual shift of $-0.01$):
\begin{equation}
\rho = -1.44 \pm 0.05  \ \ \ \ \ \ (n=11)
\label{result}
\end{equation}
for a slope $\delta_V = -2.77 \pm 0.08$ and a weighted
mean $\langle \log P \rangle=0.82 $. The adopted V-band PL relation is then 
$ \langle M_V \rangle = -2.77 \pm 0.08 \ \log P - 1.44 \pm 0.05$ or 
$ \langle M_V \rangle + 4.21 = -2.77 \ (\log P - 1)$.

\section*{Acknowledgements}
We thank L. Szabados for communicating his list of binary Cepheids and the referee
for his valuable comments.

\section*{APPENDIX A}
At present, the Hubble Space Telescope has observed Cepheids in 19 galaxies 
(see Lanoix et al. 1999b for an extensive compilation). These observations
are made in two bandpasses (V and I), so that we need a calibration of the PL relation both 
in V and I to apply a dereddening procedure (see Freedman et al. 1994, for instance) and compute the distance
moduli of these galaxies.   

With this aim in view for a future paper, we then perform the I calibration based on HIPPARCOS parallaxes
in the light of our V calibration.
The major problem is that there's no homogeneous I photometry available for each Cepheid
of the calibrating sample, and that a selection may induce a biased result. 
As a matter of fact we found I (Cousins) photometry for 174 Cepheids of the sample from Caldwell \& Coulson 
(1987). We apply to these values
a tiny correction (0.03 mag) in order to convert them into intensity averaged magnitudes.
The I magnitudes of these stars are listed in table 4 when available.
Since the selection doesn't come from a rough cut in the HIPPARCOS sample, it will not necessarily lead us
to a biased result.   
We then apply the same selection to the V sample and compute again the visual
zero-point. From these 174 cepheids we obtain:
\begin{equation}
\rho = -1.49 \pm 0.10
\end{equation}
This result is almost identical to the one obtained with the complete sample (Eq. \ref{result}), so that we conclude that this selection
implies a little bias of $0.04$ with respect to the complete sample and only $0.05$ with respect to the 
adopted final value. We will take it into account to determine 
the associated I zero-point.
 
The residuals of the I and the V PL relations are correlated so that we will apply the same procedure as we do for the
V band and obtain the same narrowing effect of the instability strip. We then need the slope of the I PL relation
as well
as the I ratio of total to selective absorption. Concerning the slope that is well determined, we choose 
$\delta_I = -3.05$ (see Gieren et al. 1998, Madore \& Freedman 1991 for instance). Let's recall that the influence
of a variation of the slope is very weak.
Concerning $R_{I}$, we choose according to Caldwell \& Coulson (1987):
\begin{equation}
R_{I} = 1.82 + 0.20 (\langle B \rangle _0- \langle V \rangle _0) + 0.02 E_{(B-V)} 
\end{equation}   
The calculus leads to :
\begin{equation}
\rho_{I} = -1.84 \pm 0.09
\end{equation}       

\begin{figure}
\epsfxsize=8.5cm
\hbox{\epsfbox{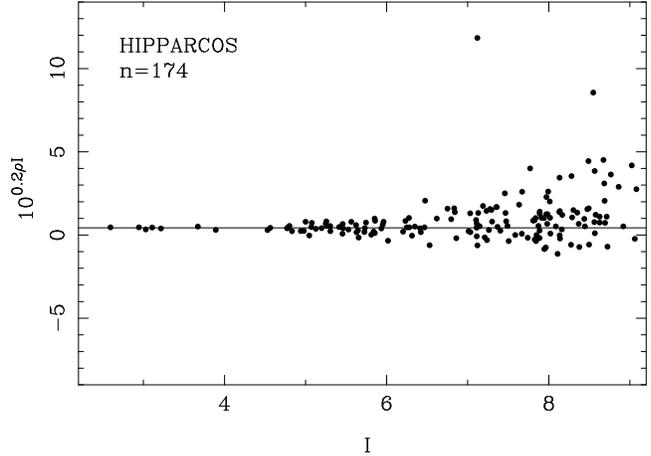}}
\caption{Position of the 174 remaining Cepheids in the I diagram zero-point vs. magnitude. The horizontal line
corresponds to $\rho_{I} = -1.84$.}
\label{rho_i}
\end{figure}         
Figure \ref{rho_i} shows that these 174 Cepheids still have an almost symetrical
distribution around a mean value, and that only faint stars with low weights
have been rejected from the sample. 
That may explain why the result is only slightly biased.

Keeping in mind that the instability strip is narrower in I than in V band, the bias du to the selection of this 
sample should be less than $0.04$ mag. Finally we adopt:
\begin{equation}
\rho_{I} = -1.81 \pm 0.09
\end{equation}
for a slope $\delta_I = -3.05$. 

\section*{APPENDIX B} 
We also investigated the effect of binarity as pointed out by Szabados (1997).
We indeed found that the dispersion of the zero-point is reduced when
only non-binary Cepheids are used. However, we interpreted this effect
by the fact that confirmed non-binary Cepheids are brighter. Actually,
using either non-binary (Evans 92) or binary (Szabados private communication)
Cepheids does not affect significantly the value of the zero-point.

\clearpage

\clearpage 
\newpage
\small
\begin{verbatim}
Table 4: The 238 Cepheids from HIPPARCOS. 
=====================================================
 Name       logPo    Pi  Sig_pi   <V>     <B>    <I>
=====================================================
 eta Aql     .856   2.78   .91   3.897   4.686  3.029
 FF Aql      .806   1.32   .72   5.372   6.128  4.531
 FM Aql      .786   2.45  1.11   8.270   9.547  6.797
 FN Aql     1.138   1.53  1.18   8.382   9.596  7.030
 SZ Aql     1.234    .20  1.10   8.599   9.988  7.032
 TT Aql     1.138    .41   .96   7.141   8.433  5.725
 V336 Aql    .864    .75  1.47   9.848  11.160  8.369
 V493 Aql    .475  -2.77  2.43  11.083  12.363
 V496 Aql    .992  -3.81  1.05   7.751   8.897  6.475
 V600 Aql    .859   1.42  1.80  10.037  11.499  8.288
 V1162 Aql   .888    .15  1.15   7.798   8.688  
 U Aql       .847   2.05   .93   6.446   7.470  5.264
 V340 Ara   1.318    .06  2.12  10.164  11.703  8.568
 AN Aur     1.013  -1.19  2.34  10.455  11.673
 BK Aur      .903    .47  1.38   9.427  10.489
 RT Aur      .571   2.09   .89   5.446   6.041  4.772
 RX Aur     1.065   1.32  1.02   7.655   8.664  6.619
 SY Aur     1.006   1.15  1.70   9.074  10.074  7.889
 Y Aur       .587   -.40  1.47   9.607  10.518
 YZ Aur     1.260   3.70  2.10  10.332  11.707
 RW Cam     1.215   -.69  2.63   8.691  10.042  7.120
 RX Cam      .898   1.14   .84   7.682   8.875  6.279
 AQ Car      .990   1.02   .81   8.851   9.779  7.889
 CN Car      .693   5.11  1.53  10.700  11.789
 CY Car      .630   -.30  1.40   9.782  10.735
 ER Car      .888   1.36   .69   6.824   7.691  5.953
 EY Car      .459   3.46  1.62  10.318  11.172
 FN Car      .661  -1.91  2.48  11.542  12.643
 FR Car     1.030    .35  1.29   9.661  10.782  8.445
 GH Car      .915    .43  1.03   9.177  10.109  8.088
 GI Car      .802   -.41  1.10   8.323   9.062  7.505 
 GX Car      .857   1.43  1.12   9.364  10.407  8.136
 GZ Car      .774   1.93  1.22  10.261  11.240  9.081
 HW Car      .964   -.71  1.06   9.163  10.218
 IT Car      .877   1.00   .82   8.097   9.087  7.111
 l Car      1.551   2.16   .47   3.724   5.023  2.593
 SX Car      .687   2.48  1.06   9.089   9.976  8.013
 U Car      1.589   -.04   .62   6.288   7.471  5.045
 UW Car      .728   -.64  1.12   9.426  10.397  8.275
 UX Car      .566    .00   .87   8.308   8.935  7.586
 UZ Car      .716   -.70  1.00   9.323  10.198  8.376
 V Car       .826    .34   .58   7.362   8.234  6.422
 VY Car     1.276   1.28  1.76   7.443   8.614  6.275
 WW Car      .670   4.23  1.39   9.743  10.633  8.675
 WZ Car     1.362   -.41  1.14   9.247  10.389  7.946
 XX Car     1.196   -.63   .95   9.322  10.376  8.108
 XY Car     1.095   -.62   .95   9.295  10.509  7.963
 XZ Car     1.221   -.30   .96   8.601   9.867  7.237
 YZ Car     1.259   1.79  1.03   8.714   9.838  7.458
 BP Cas      .797   -.60  2.04  10.920  12.470
 BY Cas      .662   -.85  3.25  10.366  11.645  
 CD Cas      .892   1.91  1.58  10.738  12.187
 CF Cas      .688  -3.20  2.16  11.136  12.310  9.752
 CH Cas     1.179    .21  1.68  10.973  12.623 
 CY Cas     1.157   2.76  3.21  11.641  13.379
 DD Cas      .992    .57  1.14   9.876  11.064  8.562
 DL Cas      .903   2.32  1.09   8.969  10.123  7.634
 DF Cas      .584   -.27  3.65  10.848  12.029  
 DW Cas      .699   1.19  1.95  11.112  12.587
 FM Cas      .764    .10  1.27   9.127  10.116  8.021 
 RS Cas      .799   2.43  1.24   9.932  11.422
 RW Cas     1.170    .69  1.68   9.117  10.213  7.910
 RY Cas     1.084    .02  1.38   9.927  11.311
 SU Cas      .440   2.31   .58   5.970   6.673  5.127
\end{verbatim}

\begin{verbatim}
Table 4: (continued)
=====================================================
 Name       logPo    Pi  Sig_pi   <V>     <B>    <I>
=====================================================   
 SW Cas      .736   1.07  1.37   9.705  10.786  8.439
 SY Cas      .610   2.73  1.49   9.868  10.860  
 SZ Cas     1.299   2.21  1.60   9.853  11.272  8.133
 UZ Cas      .629   4.37  3.64  11.338  12.448
 V636 Cas    .923   1.72   .81   7.199   8.564
 VV Cas      .793  -4.78  4.18  10.724  11.867
 VW Cas      .778  -2.12  3.61  10.697  11.942
 XY Cas      .653   -.02  1.58   9.935  11.082  
 AY Cen      .725   -.24  1.04   8.830   9.839  7.740
 AZ Cen      .660   -.20  1.04   8.636   9.289  7.887
 BB Cen      .757   3.03  1.43  10.073  11.026  9.023
 KK Cen     1.086  -1.84  2.89  11.480  12.762  9.962
 KN Cen     1.532  -1.38  2.82   9.870  11.452  7.992
 V Cen       .740    .05   .82   6.836   7.711  5.810
 V339 Cen    .976    .33  1.16   8.753   9.944  7.404
 V378 Cen    .969    .96  1.02   8.460   9.495  7.301
 V419 Cen    .898   1.72   .93   8.186   8.944  7.351
 V496 Cen    .646   1.61  1.53   9.966  11.138  8.579
 V737 Cen    .849   3.71   .84   6.719   7.718
 VW Cen     1.177  -2.02  3.63  10.245  11.590  8.766
 XX Cen     1.040   2.04   .94   7.818   8.801  6.750
 AK Cep      .859    .22  2.52  11.180  12.521 
 IR Cep      .325   1.38   .61   7.784   8.672
 CR Cep      .795   1.67  1.06   9.656  11.052  8.017
 CP Cep     1.252   1.54  1.52  10.590  12.258
 del Cep     .730   3.32   .58   3.954   4.611  3.217
 AV Cir      .486   3.40  1.09   7.439   8.349
 AX Cir      .722   3.22  1.22   5.880   6.621  5.000
 RW CMa      .758   3.12  2.16  11.096  12.321
 RY CMa      .670    .96  1.09   8.110   8.957  7.146
 RZ CMa      .629  -1.95  1.51   9.697  10.701  8.494
 SS CMa     1.092   -.37  1.75   9.915  11.127  8.497
 TV CMa      .669    .90  1.97  10.582  11.757
 TW CMa      .845   1.26  1.51   9.561  10.531  8.475
 VZ CMa      .648   1.58  1.65   9.383  10.340  8.166
 AD Cru      .806   1.87  2.32  11.051  12.330
 BG Cru      .678   1.94   .57   5.487   6.093  4.781
 R Cru       .765   1.97   .82   6.766   7.538  5.963
 S Cru       .671   1.34   .71   6.600   7.361  5.731
 SU Cru     1.109   3.93  4.73   9.796  11.548  7.672
 T Cru       .828    .86   .62   6.566   7.488  5.647
 CD Cyg     1.232    .46  1.00   8.947  10.213  7.490
 DT Cyg      .549   1.72   .62   5.774   6.312  5.197
 GH Cyg      .893   1.93  1.67   9.924  11.190
 MW Cyg      .775  -1.63  1.30   9.489  10.805  7.941
 SU Cyg      .585    .51   .77   6.859   7.434  6.203
 SZ Cyg     1.179    .86  1.09   9.432  10.909  7.825
 TX Cyg     1.168    .50  1.09   9.511  11.295  7.262
 V386 Cyg    .721   2.22  1.17   9.635  11.126  7.836
 V402 Cyg    .640   1.19  1.18   9.873  10.881  8.714
 V459 Cyg    .860    .51  1.50  10.601  12.040  8.919
 V495 Cyg    .827   -.95  1.32  10.621  12.244
 V520 Cyg    .607   1.51  1.73  10.851  12.200
 V532 Cyg    .670    .84   .94   9.086  10.122  7.872
 V538 Cyg    .787    .10  1.52  10.456  11.739
 V924 Cyg    .903    .83  1.64  10.710  11.557  9.760
 V1334 Cyg   .523    .93   .66   5.871   6.375  5.305
 VX Cyg     1.304    .88  1.43  10.069  11.773
 VY Cyg      .895   -.02  1.44   9.593  10.808  8.134
 VZ Cyg      .687   2.84  1.17   8.959   9.835  7.971
 X Cyg      1.214   1.47   .72   6.391   7.521  5.249
 bet Dor     .993   3.14   .59   3.731   4.538  2.944
 AA Gem     1.053  -2.25  2.42   9.721  10.782  8.566
 AD Gem      .578   -.18  1.60   9.857  10.551  9.061
\end{verbatim}

\begin{verbatim}
Table 4: (continued)
=====================================================
 Name       logPo    Pi  Sig_pi   <V>     <B>    <I>
=====================================================   
 DX Gem      .650  -2.58  2.49  10.746  11.682  9.622
 RZ Gem      .743   1.90  1.97  10.007  11.032  8.688
 W Gem       .898    .86  1.16   6.950   7.839  5.935
 zet Gem    1.007   2.79   .81   3.918   4.716  3.108
 BG Lac      .727   -.35  1.31   8.883   9.832  7.827
 RR Lac      .807    .94   .95   8.848   9.733  7.807
 V Lac       .697    .34   .85   8.936   9.809  7.887
 X Lac       .893    .57   .79   8.407   9.308  7.368
 Y Lac       .636  -1.53  1.21   9.146   9.877  8.305
 Z Lac      1.037   2.04   .89   8.415   9.510  7.188
 V473 Lyr    .321   1.94   .62   6.182   6.814  5.528
 AC Mon      .904    .90  1.94  10.067  11.232  8.628
 BE Mon      .432   -.28  2.12  10.578  11.712
 CV Mon      .730   3.76  2.77  10.299  11.596  8.684
 EK Mon      .597   -.77  2.69  11.048  12.243
 SV Mon     1.183  -1.18  1.14   8.219   9.267  7.130
 T Mon      1.432    .42  1.64   6.124   7.290  4.978
 TX Mon      .940    .00  2.47  10.960  12.056  9.661
 TZ Mon      .871   1.61  2.12  10.761  11.877
 V465 Mon    .434   2.28  1.88  10.379  11.141
 V508 Mon    .616  -2.42  2.28  10.518  11.416
 V526 Mon    .580   3.43  1.12   8.597   9.190
 R Mus       .876   1.69   .59   6.298   7.055  5.457 
 RT Mus      .489   1.13   .99   9.022   9.856  7.981
 S Mus       .985   2.00   .65   6.118   6.951  5.257
 UU Mus     1.066   2.85  1.27   9.781  10.931  8.489
 GU Nor      .538   4.45  2.06  10.411  11.684  8.861
 IQ Nor      .915   -.24  3.08   9.566  10.880  8.139
 RS Nor      .792   -.23  1.81  10.027  11.314
 S Nor       .989   1.19   .75   6.394   7.335  5.414
 SY Nor     1.102   2.78  1.84   9.513  10.853  
 TW Nor     1.032  -5.57  3.47  11.704  13.634  9.339
 U Nor      1.102   2.52  1.28   9.238  10.814  7.358
 BF Oph      .609   1.17  1.01   7.337   8.205  6.411
 Y Oph      1.400   1.14   .80   6.169   7.546  4.564
 CS Ori      .590   -.54  3.36  11.381  12.305
 RS Ori      .879   2.02  1.45   8.412   9.357  7.278
 AS Per      .697    .56  1.84   9.723  11.025  8.160
 AW Per      .811   2.20  1.13   7.492   8.547  6.232
 SV Per     1.046  -3.32  1.54   9.020  10.049  7.769
 SX Per      .632  -1.59  2.96  11.158  12.313
 V440 Per    .879   1.62   .83   6.282   7.155  5.303
 VX Per     1.037   1.08  1.48   9.312  10.470  7.969
 UX Per      .660  23.29  7.15  11.664  12.691
 AD Pup     1.133  -4.05  1.74   9.863  10.912
 AP Pup      .706   1.07   .64   7.371   8.209  6.467
 AQ Pup     1.479   8.85  4.03   8.791  10.214  7.119
 AT Pup      .824   1.20   .74   7.957   8.740  7.103
 BM Pup      .857   7.53 10.55  10.817  12.022
 BN Pup     1.136   4.88  1.72   9.882  11.068  8.549
 EK Pup      .571   3.54  2.34  10.664  11.480
 MY Pup      .913    .65   .52   5.677   6.308  4.941
 RS Pup     1.618    .49   .68   6.947   8.340  5.461
 VW Pup      .632  -5.65  2.83  11.365  12.430
 VZ Pup     1.365   1.49  1.47   9.621  10.783  8.280
 WW Pup      .742   2.07  1.91  10.554  11.428
 WX Pup      .951  -1.05  1.08   9.063  10.031  7.985
 WY Pup      .720    .11  2.09  10.569  11.360  9.747
 WZ Pup      .701   -.55  1.77  10.326  11.115  9.408
 X Pup      1.415   -.05  1.10   8.460   9.587  7.111
 KQ Sco     1.459    .07  2.31   9.807  11.741  7.667
 RV Sco      .783   2.54  1.13   7.040   7.995  5.857
 V482 Sco    .656   -.45  1.16   7.965   8.940  6.859
\end{verbatim}

\begin{verbatim}
Table 4: (continued)
=====================================================
 Name       logPo    Pi  Sig_pi   <V>     <B>    <I>
=====================================================   
 V500 Sco    .969   2.21  1.30   8.729  10.005  7.232
 V636 Sco    .832   -.45   .89   6.654   7.590  5.655
 V950 Sco    .529   2.46  1.04   7.302   8.077  
 CK Sct      .870   3.62  2.12  10.590  12.156
 CM Sct      .593  -3.72  2.35  11.106  12.477  9.479
 EV Sct      .643    .91  1.92  10.137  11.297  8.694
 RU Sct     1.294    .89  1.61   9.466  11.111  7.474
 SS Sct      .565  -1.07  1.17   8.211   9.155  7.110
 TY Sct     1.043   4.02  2.27  10.831  12.488
 X Sct       .623    .97  1.46  10.006  11.146  8.628
 Y Sct      1.015    .00  1.69   9.628  11.167  7.849
 Z Sct      1.111   1.14  1.66   9.600  10.930  8.131
 CR Ser      .724  -3.04  2.08  10.842  12.486
 S Sge       .923    .76   .73   5.622   6.427  4.832
 AP Sgr      .704   -.95   .92   6.955   7.762  6.018
 AY Sgr      .818   -.99  2.28  10.549  12.006
 BB Sgr      .822    .61   .99   6.947   7.934  5.840
 U Sgr       .829    .27   .92   6.695   7.782  5.455
 V350 Sgr    .712   -.10  1.05   7.483   8.388  6.314
 W Sgr       .880   1.57   .93   4.668   5.414  3.892
 WZ Sgr     1.340   -.75  1.76   8.030   9.422  6.530
 X Sgr       .846   3.03   .94   4.549   5.288  3.671
 Y Sgr       .761   2.52   .93   5.744   6.600  4.801
 YZ Sgr      .980    .87  1.03   7.358   8.390  6.248
 EU Tau      .473    .86  1.38   8.093   8.757
 ST Tau      .606   3.15  1.17   8.217   9.064
 SZ Tau      .651   3.12   .82   6.531   7.375  5.564
 S TrA       .801   1.59   .72   6.397   7.149  5.623
 R TrA       .530    .43   .71   6.660   7.382  5.853
 alf UMi     .754   7.56   .48   1.982   2.580  1.393
 AE Vel      .853   -.64  1.33  10.262  11.505  8.723
 AH Vel      .782   2.23   .55   5.695   6.274  5.078
 BG Vel      .840   1.33   .65   7.635   8.810  6.348
 DR Vel     1.049   -.45  1.07   9.520  11.038  7.842
 RY Vel     1.449  -1.15   .83   8.397   9.749  6.841
 RZ Vel     1.310   1.35   .63   7.079   8.199  5.852
 ST Vel      .768  -1.62   .99   9.704  10.899  8.351
 SV Vel     1.149  -1.27   .97   8.524   9.758  7.466
 SW Vel     1.370   1.30   .90   8.120   9.282  6.834
 SX Vel      .980   1.54   .79   8.251   9.139  7.293
 T Vel       .666    .48   .72   8.024   8.946  7.010
 XX Vel      .844   1.14  1.50  10.654  11.816
 BR Vul      .716  -2.80  1.70  10.687  12.161  
 SV Vul     1.653    .79   .74   7.220   8.662  5.746
 T Vul       .647   1.95   .60   5.754   6.389  5.071
 U Vul       .903    .59   .77   7.128   8.403  5.630
 X Vul       .801   -.33  1.10   8.849  10.238  7.210
\end{verbatim}
 
\end{document}